%
%
\input harvmac
\newcount\yearltd\yearltd=\year\advance\yearltd by 0

\noblackbox

\input epsf

\def\mod{\rm \, mod \, }
\newcount\figno
\figno=0
\def\fig#1#2#3{
\par\begingroup\parindent=0pt\leftskip=1cm\rightskip=1cm\parindent=0pt
\baselineskip=11pt
\global\advance\figno by 1
\midinsert
\epsfxsize=#3
\centerline{\epsfbox{#2}}
\vskip 12pt
{\bf Fig.\ \the\figno: } #1\par
\endinsert\endgroup\par
}
\def\figlabel#1{\xdef#1{\the\figno}}
\def\encadremath#1{\vbox{\hrule\hbox{\vrule\kern8pt\vbox{\kern8pt
\hbox{$\displaystyle #1$}\kern8pt}
\kern8pt\vrule}\hrule}}

\def\inbar{\vrule height1.5ex width.4pt depth0pt}

 \def\d{{\delta}}

 \def\frac#1#2{{#1\over #2}}

 \def\CN{{\cal N}}

\def\e{{\epsilon}}

\def\IR{\relax{\rm I\kern-.18em R}}
\def\IC{\relax\hbox{$\inbar\kern-.3em{\rm C}$}}
\def\IZ{\relax\ifmmode\hbox{Z\kern-.4em Z}\else{Z\kern-.4em Z}\fi}


\lref\SundborgUE{
B.~Sundborg,
``The Hagedorn transition, deconfinement and $N = 4$ SYM theory,''
Nucl.\ Phys.\ B {\bf 573}, 349 (2000)
[arXiv:hep-th/9908001].
}
\lref\apostol{
Tom Apostol
}

\lref\MinwallaKA{
S.~Minwalla,
Adv.\ Theor.\ Math.\ Phys.\  {\bf 2}, 781 (1998)
[arXiv:hep-th/9712074].
}

\lref\AharonyTI{
O.~Aharony, S.~S.~Gubser, J.~M.~Maldacena, H.~Ooguri and Y.~Oz,
Phys.\ Rept.\  {\bf 323}, 183 (2000)
[arXiv:hep-th/9905111].
}

\lref\MaldacenaRE{
J.~M.~Maldacena,
Adv.\ Theor.\ Math.\ Phys.\  {\bf 2}, 231 (1998)
[Int.\ J.\ Theor.\ Phys.\  {\bf 38}, 1113 (1999)]
[arXiv:hep-th/9711200].
}

\lref\AharonySX{
O.~Aharony, J.~Marsano, S.~Minwalla, K.~Papadodimas and M.~Van Raamsdonk,
arXiv:hep-th/0310285.
}

\lref\MinwallaKA{
S.~Minwalla,
Adv.\ Theor.\ Math.\ Phys.\  {\bf 2}, 781 (1998)
[arXiv:hep-th/9712074].
}

\lref\DolanZH{
F.~A.~Dolan and H.~Osborn,
Annals Phys.\  {\bf 307}, 41 (2003)
[arXiv:hep-th/0209056].
}

\lref\BeisertDI{
N.~Beisert, M.~Bianchi, J.~F.~Morales and H.~Samtleben,
JHEP {\bf 0407}, 058 (2004)
[arXiv:hep-th/0405057].
}

\lref\BianchiWX{
M.~Bianchi, J.~F.~Morales and H.~Samtleben,
JHEP {\bf 0307}, 062 (2003)
[arXiv:hep-th/0305052].
}

\lref\BianchiXI{
M.~Bianchi,
arXiv:hep-th/0409304.
}

\lref\BianchiWW{
M.~Bianchi,
arXiv:hep-th/0409292.
}

\lref\BeisertTE{
N.~Beisert, M.~Bianchi, J.~F.~Morales and H.~Samtleben,
JHEP {\bf 0402}, 001 (2004)
[arXiv:hep-th/0310292].
}

\lref\DobrevQV{
V.~K.~Dobrev and V.~B.~Petkova,
Phys.\ Lett.\ B {\bf 162}, 127 (1985).
}

\lref\DobrevVH{
  V.~K.~Dobrev and V.~B.~Petkova,
  Lett.\ Math.\ Phys.\  {\bf 9}, 287 (1985).
}

\lref\DobrevQZ{
  V.~K.~Dobrev and V.~B.~Petkova,
  Fortsch.\ Phys.\  {\bf 35}, 537 (1987).
}

\lref\DobrevSP{
V.~K.~Dobrev and V.~B.~Petkova,   
Proceedings, eds. A.O. Barut and H.D. Doebner, Lecture Notes in
Physics, Vol. 261 (Springer-Verlag, Berlin, 1986) pp. 300-308.
}

\lref\DobrevTK{
  V.~K.~Dobrev,
  arXiv:hep-th/0406154.
}

\lref\MackJE{
G.~Mack,
Commun.\ Math.\ Phys.\  {\bf 55}, 1 (1977).
}

\lref\BarsEP{
I.~Bars and M.~Gunaydin,
Commun.\ Math.\ Phys.\  {\bf 91}, 31 (1983).
}

\lref\BerensteinJQ{
D.~Berenstein, J.~M.~Maldacena and H.~Nastase,
JHEP {\bf 0204}, 013 (2002)
[arXiv:hep-th/0202021].
}

\lref\BeisertFV{
N.~Beisert, G.~Ferretti, R.~Heise and K.~Zarembo,
arXiv:hep-th/0412029.
}

\lref\BeisertRY{
N.~Beisert,
Phys.\ Rept.\  {\bf 405}, 1 (2005)
[arXiv:hep-th/0407277].
}

\vskip 0.8cm

\Title{\vbox{\baselineskip12pt\hbox{hep-th/0501063}}}{\vbox{\centerline{The Spectrum of Yang Mills on a Sphere}}}

\centerline{Alexander Barabanschikov$^{a}$, Lars Grant$^{b,c}$,Lisa L Huang$^{b}$ and Suvrat Raju$^{b,c}$}
\smallskip
\centerline{\sl $^{a}$ Department of Physics, NorthEastern University,Boston, MA 02115, USA}
\centerline{\sl $^{b}$ Jefferson Physical Laboratory,Harvard University,Cambridge, MA 02138, USA}
\centerline{\sl $^{c}$ Tata Institute of Fundamental Research, Mumbai 400005, India}

\vskip .6in 
\centerline{\bf Abstract}{ In this note, we determine the representation content of the free, large $N$, $SU(N)$ Yang Mills theory on a sphere by decomposing its thermal partition function into characters of the irreducible representations of the conformal group SO(4,2). 
We
also discuss the generalization of this procedure to finding the
representation content of $\CN=4$ Super Yang Mills. }\vskip .3in

\smallskip
\Date{}


\newsec{Introduction}

Almost thirty years ago t'Hooft, Polyakov, Migdal and Wilson
suggested that large $N$ Yang Mills theory could be recast as a
string theory. Electric flux tubes of the confining gauge theory
were expected to map to dual fundamental strings. This picture
seemed crucially tied to confinement and strongly coupled gauge
dynamics; in particular Yang Mills perturbation theory was not
expected to capture the stringy behavior of Yang Mills theory.

Some of these expectations were modified after one
gauge-string duality  was understood in detail via the
celebrated AdS/CFT conjecture \refs{\MaldacenaRE, \AharonyTI}. According to the Maldacena
conjecture, maximally supersymmetric $SU(N)$ Yang Mills theory on an
$S^3$ is dual to type IIB theory on $AdS_5\times S^5$. $\CN=4$ Yang
Mills theory is a conformal (rather than confining) gauge theory,
further it possesses a tunable coupling constant $\lambda$ which may
smoothly be taken to zero. Nonetheless, the AdS/CFT conjecture
asserts that this gauge theory is dual to a string theory at all
values of $\lambda$ including $\lambda=0$.

It follows that stringy aspects of the $\CN=4$ theory must be
visible even at arbitrarily weak coupling. Evidence has been
accumulating over recent years that this is indeed the case. First,
the spectrum of weakly coupled $\CN=4$ Yang Mills theory on an $S^3$
displays a Hagedorn like growth in its density of states
\refs{\AharonySX, \SundborgUE} (a distinctly stringy feature).
Second, perturbative contributions to the anomalous dimensions of
certain long single trace operators are effectively encoded in a 1+1
dimensional field theory \BerensteinJQ\ . For recent developments in this story, please see \BeisertRY\ and references therein.

With these lessons in mind, it is natural to wonder if the stringy
spectrum of the simplest of all large $N$ Yang Mills theories - pure
$SU(N)$ Yang Mills - may also be computed (in some regime of
parameters) in perturbation theory. Naively it seems that
asymptotically free pure Yang Mills lacks a dimensionless parameter
in which one might attempt a perturbative expansion. However Yang
Mills theory on an $S^3$ of radius $R$  does have a dimensionless
coupling $\lambda = \Lambda R$, where $\Lambda$ is the dynamically
generated scale. When $\lambda$ is taken to zero, the spectrum of
pure Yang Mills may reliably be studied in perturbation theory, and
turns out to display the same Hagedorn growth with energy as its
supersymmetric counterpart \refs{\AharonySX, \SundborgUE}. It thus
seems plausible that, like its supersymmetric counterpart,  pure
Yang Mills theory on an $S^3$ has a dual string description at all
values of $\lambda$.

In this note we will compute the spectrum of particles of the as
yet unknown string dual to pure Yang Mills on a sphere at
$\lambda=0$. We do this by decomposing the partition function of our
theory at $\lambda=0$ (computed in \refs{\AharonySX, \SundborgUE} )
into a sum over characters of the conformal group (the conformal
group is a good symmetry of pure Yang Mills theory precisely at
$\lambda=0$). According to the AdS/CFT dictionary, representations
of the conformal group are in one to one correspondence with
particles of the dual string theory; consequently our decomposition
determines particle spectrum of interest.

The plan of the rest of this note is as follows. In sections 1-4 we
review the conformal algebra and calculate the characters of the
representations. Since this involves some subtleties, in Section 5, we verify the
calculation using the oscillator construction proposed in \BarsEP\ . The
integral we need to perform to obtain the  character decomposition has
poles. Section 6 discusses the appropriate treatment
for these,
while the actual calculation is performed in Section 7. We conclude
with some comments on the supersymmetric case. In the Appendix
we 
prove that the multiplicities obtained in Section 7 are integers.

While we were in the process of writing this paper, we received \BeisertFV\ that has overlaps with Sections 3,4.

\newsec{Conformal Algebra}

Adding dilatations and special conformal transformations to a set of Lorentz generators in 4 dimensions gives the conformal algebra.
\eqn\conformalalg{
\eqalign{
[D,P_{\mu}] &= -i P_{\mu}, \cr
[D,K_{\mu}] &= i K_{\mu}, \cr
[K_{\mu}, P_{\nu}] &= 2 i (\eta_{\mu \nu} D + M_{\mu \nu}), \cr
[M_{\mu \nu}, P_{\rho}] &= i(\eta_{\mu \rho} P_{\nu} - \eta_{\nu \rho} P_{\mu}) ,\cr
[M_{\mu \nu},K_{\rho}] &= i(\eta_{\mu \rho} K_{\nu} - \eta_{\nu \rho} K_{\mu}), \cr
[M_{\mu \nu}, M_{\rho \sigma}] &= i\left(\eta_{\mu \rho} M_{\nu \sigma} + \eta_{\nu \sigma} M_{\mu \rho} -
\eta_{\mu \sigma}  M_{\nu \rho} - \eta_{\nu \sigma} M_{\mu \rho} \right) .\cr
}
}
We are interested in unitary representations of this algebra, where these generators are hermitian.
However, it is convenient for the purposes of constructing the representations of this algebra, to choose
a basis of generators which satisfies the euclidean conformal algebra \MinwallaKA\ and in which the generators
are no longer all hermitian.  The generators (some $D'$, $P'^{\mu}$, etc.) in this new basis will satisfy
the same algebra as above with $\eta_{\mu \nu} \to \delta_{\mu \nu}$.  The hermiticity properties of the generators in
this basis are:
$$
\eqalign{
M'^\dagger &= M', \cr
D'^\dagger &= - D', \cr
P'^\dagger &= K', \cr
K'^\dagger &= P'. \cr
}
$$
From now on, we will use this new set of generators and drop the primes for clarity.

We can extract two sets of $SU(2)$ generators from the Lorentz generators $M_{\mu \nu}$.
We define:
\eqn\jdef{
\eqalign{
J_1^z&=1/2(M_{1 2}+M_{0 3}), \cr
J_2^z&=1/2(M_{1 2}-M_{0 3}), \cr
J_1^+&=1/2(M_{2 3}+M_{0 1} + i(M_{0 2} - M_{1 3})), \cr
J_2^+&=1/2(M_{2 3}-M_{0 1} - i(M_{1 3} + M_{0 2})), \cr
J_1^-&=J_1^{+ \dagger}, \quad J_2^-=J_2^{+ \dagger}.
}
}
We will also choose to use an hermitian operator $D''=iD$ for convenience. We note that the set of
generators $M= \{D'',J_1^z, J_1^+,J_1^-,J_2^z, J_2^+,J_2^-\}$ generate the maximal compact subgroup
$SO(2)\times SO(4) \subset SO(4,2)$ of the conformal group. We can divide the generators into
three sets :
$G_0 = \{D,J_1^{z}, J_2^{z} \}, G_{+} = \{J_1^{+},J_2^{+},P_{\mu}\}, G_{-} = \{ K_{\mu},J_1^{-}, J_2^{-}\}.$
With this division, the Lie algebra above has the property that:
$$
\eqalign{
[g_0, g_{+}] &= g_{+}^{'}, \cr
[g_0,g_0^{'}] &= g_{0}^{''}, \cr
[g_{+},g_{-}] &= g_{0}, \cr
[g_0,g_{-}] &= g_{-}^{'}.
}
$$
where anything with a subscript $0$ belongs to linear combinations of operators in $G_0$ and similarly symbols with subscripts $+(-)$
belong to linear combinations of operators in $G_{+}(G_{-})$. These relations make it clear
that $G_+$ and $G_-$ act like raising and lowering operators on the charges $G_0$.
The operators in $G_0$ commute and we will use these as Cartan generators for the algebra.

It will be convenient to choose linear combinations of the operators in $G_+$ and $G_-$ that diagonalize $G_0$.  These combinations are:
\eqn\Pdef{
\eqalign{
P_w=P_1+i P_2, \cr
P_{\overline{w}}=P_1- i P_2 ,\cr
P_z=P_3+i P_4, \cr
P_{\bar{z}}=P_3-i P_4.
}
}
These generators all have well defined weights under the Cartan generators $G_0$.
\eqn\weights{
\matrix{
& D & J_1 & J_2 \cr
J_1^+ & 0 & 1 & 0 \cr
J_2^+ & 0 & 0 & 1 \cr
P_w & 1 & {1 \over 2} & {1 \over 2} \cr
P_{\bar{w}} & 1 & -{1 \over 2} & -{1 \over 2} \cr
P_z & 1 & -{1 \over 2} & {1 \over 2} \cr
P_{\bar{z}} & 1 & {1 \over 2} & -{1 \over 2}
}}
\newsec{Representations of the Conformal Group}

Any irreducible representation of the conformal group can be written as some direct sum of representations of the compact
subgroup $SO(4)\times SO(2)$:
\eqn\reduction{
R_{SO(4,2)}= \sum_i \bigoplus R_{comp}^i.
}
We are ultimately interested in the occurrence of these representations in the partition function of the conformal Yang-Mills gauge theory quantized on $S^3 \times R$;  the hamiltonian of the theory is the dilatation operator $D$.  The spectrum of this theory is bounded below and therefore we will be interested in representations of the
conformal algebra where the values of the charge $D$ are bounded below.  Then there must be some term, $R_{comp}^{\lambda} $
in the above sum that has the lowest dimension. This term has a highest weight state $|\lambda>$ with weights $\lambda=(d,j_1,j_2)$.
The $K^{\mu}$ operators necessarily annihilate all the states in $R_{comp}^{\lambda} $ because the $K^{\mu}$ have negative weight under
the operator $D$.  If we consider the operation of the $P^{\mu}$ on this set of states, we generate a whole representation of the conformal
algebra with states:
\eqn\verma{
[\lambda]^* = R_{SO(4,2)}^{\lambda} = \sum_{n_w n_{\bar{w}} n_z n_{\bar{z}}} P_w^{n_w}P_{\bar{w}}^{n_{\bar{w}}}P_z^{n_z}P_{\bar{z}}^{n_{\bar{z}}} \times R_{comp}^{\lambda}.
}
We will denote this set of states by $[d,j_1,j_2]^*$. A careful analysis \MackJE\ shows that, barring the trivial case, this representation is unitary if one of the
following conditions holds on the highest weight state $|\lambda>$:
\eqn\unitarity{
\eqalign{
(i) &\quad d \ge j_1 + j_2 +2 \quad j_1 \ne 0 \quad j_2 \ne 0, \cr
(ii) &\quad d \ge j_1 + j_2 + 1 \quad j_1 j_2 = 0.
}}
In the case where equality does not hold in these unitarity conditions, the representation is called {\it long} and all the states produced by
the operation of the $P^{\mu}$ are non-zero.

If equality holds in one of the conditions, then the representation will be a truncated {\it short} representation in which some of
the states listed in \verma\ are $0$. A unitary representation is one where we can define a positive definite norm.  To find the states that should
be absent in a short representation, one can assume that the states in $R_{comp}^{\lambda} $ are normalized in the standard way \MinwallaKA\ .  Calculating
the norm of the states $P^{\mu}|\lambda>$ will show that when equality holds in the unitarity conditions above, some of these states,
say $a_{\mu}P^{\mu}|\lambda>$ have norm $0$. This should be interpreted as meaning that this state is $0$ so that the operator $a_{\mu}P^{\mu}$
annihilates $|\lambda>$.  The descendants of $a_{\mu}P^{\mu}|\lambda>$ then also do not occur in the representation. This last statement
needs some care as we will see.

We will list here the possible types of short representations:

\item{$\bullet$} In the generic short representation, $j_1 \neq 0, j_2 \neq 0, d = j_1+j_2+2 $ the states of norm $0$ occur at level 1.
The state $|d+1,j_1-{1 \over 2}, j_2-{1 \over 2}>$ is {\it not} found in the representation and its descendants are also not to be found in the representation.
The set of all descendants of the state $\lambda'=|d+1,j_1-{1 \over 2}, j_2-{1 \over 2}>$ is the same as $[ d+1,j_1-{1 \over 2}, j_2-{1 \over 2}]^*$, so that
we may write the generic short, irreducible representation as:
\eqn\genericshort{
[d,j_1,j_2]=[d,j_1,j_2]^*-[d+1,j_1-{1 \over 2}, j_2-{1 \over 2}]^*.
}
\item{$\bullet$}In the case $j_1 = j_2 = 0$, $d=1$, the state $|3,0,0>$ is not found.  All its descendants are also absent, so we may write the irreducible representation as
\eqn\scalarshort{
[1,0,0]=[1,0,0]^*-[3,0,0]^*.
}
\item{$\bullet$}In the case $j_1 > 0, j_2=0$, the state $|d+1, j_1-{1 \over 2}, {1 \over 2} >$ is absent.  Note that the weights of this state satisfy the unitarity bound $(i)$ in
\unitarity.  When we delete the states $[d+1, j_1-{1 \over 2}, {1 \over 2} ]^*$, we must delete it as a {\it short} representation, ie. we
must not delete the states that do not occur in the short rep $[d+1, j_1-{1 \over 2}, {1 \over 2} ]$. We will do a calculation below, using an oscillator
representation, showing that this is the correct way to remove the zero norm states in this case. We will have
\eqn\funkyshort{
[d, j_1,0]= [d,j_1,0]^*-[d+1,j_1- {1 \over 2}, {1 \over 2} ]^* + [d+2,j_1-1,0]^*.
}

\newsec{Characters}

The characters for these representations are now easy to compute.  First we compute a character of the set of states $[d,j_1,j_2]^*$. We will denote
the character of this set of states by a bar on $\chi$:
\eqn\character{\eqalign{
\bar{\chi}_{[d,j_1,j_2]} &={Tr}_{[d,j_1,j_2]^*} \exp[i D \theta + i J_1^z \theta_1 + i J_2^z \theta_2] \cr
&= \sum_{\matrix{n_k \ge 0 \cr |m_1| < j_1 \cr|m_2| < j_2} } <adjoint|
e^{i D \theta + i J_1^z \theta_1 + i J_2^z \theta_2} P_{w}^{n_1} P_{\bar{w}}^{n_2} P_{z}^{n_3}
P_{\bar{z}}^{n_4} |d,m_1,m_2> \cr
&= {\chi_{j_1}^{SU(2)}\chi_{j_2}^{SU(2)}e^{i d \theta} \over \prod_{j=1}^4 (1 - \exp[i \vec{\alpha_j} \cdot \vec{\theta} ])} \cr
}}

where $\vec{\theta}=(\theta,\theta_1,\theta_2)$, and $\alpha_j$ runs over the 4 generators $P_w,P_{\bar{w}},P_z,P_{\bar{z}}$ and
refers to their weights taken from the table \weights\ , ie $\alpha_1=(1,1/2,1/2)$.

The characters of the possible representations are given by:
\eqn\characters{
\matrix{
(i) & {\rm Long} & d > j_1+j_2+2 & \chi_{[d,j_1,j_2]}=\bar{\chi}_{[d,j_1,j_2]} \cr
(ii) & {\rm Short} & j_1=j_2=0 \quad d=1 & \chi_{[1,0,0]}=\bar{\chi}_{[1,0,0]}-\bar{\chi}_{[3,0,0]} \cr
(iii) & {\rm Short} & j_1>0 \quad j_2=0 \quad d=j_1+1 & \chi_{[d,j_1,0]}=\bar{\chi}_{[d,j_1,0]}-\bar{\chi}_{[d+1,j_1-1/2,1/2]}+\bar{\chi}_{[d+2,j_1-1,0]} \cr
(iv) & {\rm Short} & j_1>0 \quad j_2 > 0 \quad d = j_1 + j_2 + 2 & \chi_{[d,j_1,j_2]} = \bar{\chi}_{[d,j_1,j_2]} - \bar{\chi}_{[d+1,j_1 - 1/2, j_2 - 1/2]}
}}
We will note shortly that these characters are {\it not} orthogonal.  Nevertheless, they can be used to decompose the spectrum of the Conformal Yang Mills theory
we are interested in.

\newsec{Oscillator Construction}

Here we will discuss an oscillator construction \BarsEP\ for the SO(4,2) algebra and use it to confirm the character of the short representations $j_2=0$ and $d=j_1+1$.  The
SO(4,2) algebra may be represented by 8 bosonic oscillators $a^I, b^J,a_I$ and $b_J$ ($I,J=1,2$) having the following commutation relations:
\eqn\commutations{
[a_I,a^J]=\delta_I^J \quad [b_P,b^Q]=\delta_P^Q.
}
The generators of the $SO(4,2)$ group are represented as:
\eqn\oscgenerators{\eqalign{
J_1^i=1/2(\sigma^i)_I^J[a^I a_J - 1/2 \delta^I_J a^K a_K], \quad J_2^i=1/2(\bar{\sigma}^i)_P^Q[b^P b_Q - 1/2 \delta^P_Q b^R b_R] \cr
D=1/2(N_a + N_b +2),\quad P^{I J}= a^I b^J, \quad K_{I J}= a_I b_J.
}}
We note that a state constructed out of oscillators acting on a vacuum satisfying $a_I|0>=b_J|0>=0$ has weights $(1/2(N_a+N_b+2), 1/2(n_{a_1}-n_{a_2}),1/2(n_{b_1}-n_{b_2}))$
under $D, J_1^z, J_2^z$ ($n_{a_1}$ is the number of $a^1$ operators used to create the state and $N_a=n_{a^1}+n_{a^2}$).
The unitarity constraints are built into this representation, so we may calculate with it without worrying about states that have norm zero.  For example, we may
compute the ``blind'' partition function of the short representation $|\lambda>=(j_1+1,j_1,0)$. We first choose a state with the right weights to act as the primary:
\eqn\osprimary{
(a^2)^{2 j_1} |0>.
}
Now we can easily generate from this state, a representation of the maximal compact subgroup $SO(4) \times SO(2)$:
\eqn\oscomp{
a^{I_1}a^{I_2}a^{I_3}\ldots a^{I_{2 j_1}}|0>.
}
There are $2 j_1 +1$ states here, all with dimension $D=j_1+1$ as we expect. Now we operate with all possible $P^{\mu}$:
\eqn\shortos{\eqalign{
Z_{[j_1+1,j_1,0]}=\sum_{n_1,n_2,m_1,m_2} \sum_{I_k} <adjoint| x^{D} a_1^{n_1}a_2^{n_2}b_1^{m_1}b_2^{m_2}a^{I_1}a^{I_2}a^{I_3}\ldots a^{I_{2 j_1}}|0> \cr
=\sum_{N=0}^{\infty} x^{N+j_1}N(N+2 j_1) \cr
={x^{j+1}(2j_1+1 - 4 j_1 x + (2j_1-1) x^2) \over (1-x)^4}.
}}
This agrees with the result in \characters\ .  In the second line, we have used the fact that $n_1+n_2=m_1+m_2$ and that the number of $a$s in \oscomp\ is $2 j_1$. This calculation can easily be repeated with chemical potentials added for the angular momenta.

\newsec{Character Decomposition}
Character decomposition integrals are evaluated over the Haar measure of the group in question, in this case $SO(4,2)$.
We can reduce these integrals to integrals over the maximal torus of the maximal compact subgroup
$SU(2) \times SU(2) \times SO(2)$ using the Weyl integration formula
\eqn\weyl{
\int_G f(g) d\mu_G ={1 \over |W|} \int_T f(t) \prod_{\alpha \in R}   (1-\exp(\alpha(t))) d\mu_T.
}
where $f(g)$ is a function satisfying $f(h g h^{-1})=f(g)$ so that it only depends on the conjugacy class of $g$, and $d\mu_G$ and $d\mu_T$ are the Haar measures on the
group $G$ and the maximal torus $T$.  $\alpha \in R$ means the product is over the
 roots of $SO(4,2)$.  Each root corresponds to a generator in table \weights\ , for example the factor corresponding to $K_w$ is
$(1-\exp(-i(\theta + {\theta_1+\theta_2 \over 2})))$.  The constant $|W|$ is the order of the Weyl group in the compact case.  In this non-compact case, it will diverge. We nevertheless obtain a useful orthogonality relation below where this constant is not relevant. An integral of characters over the group G becomes:

\eqn\integral{\eqalign{
\int_G \chi_{[d,j_1,j_2]}^* \chi_{[d',j_1',j_2']} d\mu_G = & {1 \over |W|} \int_0^{2 \pi} \int_0^{4 \pi} \int_0^{4 \pi}
\chi_{[d,j_1,j_2]}^*(\theta,\theta_1, \theta_2)\chi_{[d',j_1',j_2']}(\theta,\theta_1, \theta_2) \cr
& \prod_{\alpha \in R}(1-\exp(i \vec{\alpha} \cdot \vec{\theta})) {d\theta \over 2 \pi}{d\theta_1 \over 4 \pi}{d\theta_2 \over 4 \pi}
.}}
While the characters of the non-compact group $SO(4,2)$  are not orthogonal, the characters of the sets of states $[d,j_1,j_2]^{*}$ can easily be shown to explicitly satisfy the following orthogonality relation:
\eqn\orth{
{1 \over 4} \int \bar{\chi}_{[d,j_1,j_2]}^* \bar{\chi}_{[d',j_1',j_2']} \prod_{\alpha \in R}(1-\exp(i \vec{\alpha} \cdot \vec{\theta})))d\mu_T =
\delta_{d,d'} \delta_{j_1,j_1'} \delta_{j_2,j_2'}.
}
This orthogonality is enough for us to decompose the partition function of YM into representations of the conformal group.

In the case of non-compact groups, character decomposition integrals involve some subtleties. Written naively, these integrals have poles. To learn how to deal with these poles, consider the representation $[1,0,0] \times [1,0,0]$.  The decomposition of this tensor product by characters will involve integrals like
\eqn\e{\eqalign{
&\int (\bar{\chi}_{[1,0,0]})^2  \bar{\chi}^*_{[d,j_1,j_2]} d\mu_G =\cr &\int {(\cos j_1 \theta_1 -\cos (j_1 + 1) \theta_1)(\cos j_2 \theta_2 - \cos (j_2 + 1) \theta_2) \exp[-i d \theta] \exp[2 i \theta]  \over \prod_{\alpha \in P} (1 - e^{i \vec{\alpha}\cdot \vec{\theta}})}   {d\theta \over 2 \pi}{d\theta_1 \over 4 \pi}{d\theta_2 \over 4 \pi},
}}
where $\alpha \in P$ means product only over the 4 roots corresponding to momentum generators $P_i$ as in \character\ .
It is clear that this integral has singularities along the contour of integration. To resolve this, we deform the contour {\it inwards}.This is equivalent to {\it ignoring} the contribution from the boundaries.

To see why this is justified, expand
\eqn\f{
\prod{1 \over 1 - x q_i} = \sum x^n q_i^n .
}
We have introduced new notation here.  $x=e^{i\theta}$, $q_i=e^{i{\pm \theta_1 \pm \theta_2  \over 2}}$ for $i=1,2,3,4$.
Now recalling that $x$ measures the scaling dimension or the energy, we see that that we should add a small imaginary part to $\theta$ which is
equivalent to inserting an energy cutoff in the integral.

With this pole prescription, the decomposition yields:
\eqn\charonedecomposed{
\chi_{[1,0,0] * [1,0,0]} = \sum_{d=2}^{\infty}\chi_{[d,{d - 2 \over 2},{d-2 \over 2}]}.
}
These representations are generically short (barring $[2,0,0]$).

We can count the operators in our theory manually to check this result. Using two scalar field representations, the primary operators in the tensor product at the first few levels are: 
\eqn\ops{\eqalign{
&\phi_1 \phi_2 \qquad [2,0,0], \cr
&\phi_2 \partial_{\mu} \phi_1  \qquad [3,{1 \over 2},{1 \over 2}], \cr
&\partial_{{\mu}} \phi_1 \partial_{\nu} \phi_2 \qquad [4,1,1]. \cr
}}
which agrees with the decomposition.  We will use this same pole prescription in performing the decomposition of the YM theory.

\newsec{The Integral}
The single trace partition function of Free Yang Mills on a sphere was calculated in \refs{\AharonySX, \SundborgUE}. The result was written as
\eqn\singletrace{
Z[\theta,\theta_1,\theta_2] = -\sum {\phi(k) \over k} \ln(1 - z(k \theta,k \theta_1,k \theta_2)),
}
where the 'single particle' partition function, z is given by:
\eqn\singleparticle{
z = 1 + {(x - x^3) \sum_i q_i + x^4 - 1 \over \prod_i (1 - x q_i)}.
}
We need to decompose this expression as a sum of characters of the conformal group. First,
\eqn\singlefactored{
1 - z = {(1 - x^2)(x^2 - (\sum q_i) x + 1) \over \prod_i (1 - x q_i)}.
}
So, the logarithm in \singletrace\ will separate the factors here into terms which we will integrate one at a time.
Also, we have explicitly
\eqn\character{
\bar{\chi}_{d,j_1,j_2} = {\sin[(j_1+1/2) \theta_1] \over \sin[{\theta_1 \over 2}]}{ \sin[(j_2+1/2) \theta_2] \over \sin[{\theta_2 \over 2}]} \exp[i d \theta] \prod_{i}{ 1 \over (1 - x q_i)}.
}
The measure of integration is:
\eqn\measure{
dM = 4 \sin^2[{\theta_1 \over 2}] \sin^2[{\theta_2 \over 2}] \prod_{i}(x - 1/q_i)(x - q_i) {d\,\theta \over 2 \pi}  {d\, \theta_1 \over 4 \pi}{d \,\theta_2 \over 4 \pi}.
}
Note that the $\theta_1, \theta_2$ integrals go over ${0, 4\pi}$.

We will evaluate
\eqn\integral{
\int dM Z[\theta,\theta_1,\theta_2] \bar{\chi}^*_{[d,j_1,j_2]}.
}
Half of the measure cancels with the denominator of the character. The remaining part of the measure may be written as
\eqn\simplify{
4 \sin {\theta_1 \over 2} \sin {\theta_2 \over 2} \prod_{i} (x - q_i) = 4 \sin {\theta_1 \over 2} \sin {\theta_2 \over 2}\{ (x^4 + 1) - 4 \cos{ \theta_1 \over 2}
\cos {\theta_2 \over 2} (x^3 + x) + 2 x^2(\cos \theta_1 + \cos \theta_2 + 1)\}.
}
We will do the integral over the 4 linear factors in the denominator of \singlefactored\ first. The contribution from the partition function $Z$ is
\eqn\partdencontrib{
-\sum_{k,i} {\phi(k) \over k} \log{1 \over 1 - x^k q_i^k} = -\sum_{k,i,n} \phi(k) {x^{k n} q_i^{k n} \over k n} = -\sum_{k,n}  \phi(k) {4 \cos {k n \theta_1 \over 2}
\cos {k n \theta_2 \over 2} x^{k n} \over k n}.
}
The integration over $\theta$ picks out coefficients of $x^d$ in the product of \partdencontrib\ and the measure \simplify\ . The coefficient of $x^d$ in \partdencontrib\ is
\eqn\polecollection{
c(d) = -\sum_{k | d} {\phi(k) \over d} 4 \cos {d \theta_1 \over 2} \cos {d \theta_2 \over 2} = - 4 \cos {d \theta_1 \over 2} \cos {d \theta_2 \over 2}.
}
Hence, we need to deal with the integral
\eqn\long{
\eqalign{
\int
\left[c(d) + c(d-4) - 4 \cos {\theta_1 \over 2} \cos {\theta_2 \over 2}(c(d-1) + c(d-3)) + 2(\cos \theta_1 + \cos \theta_2 + 1)c(d-2)\right] \cr (\cos j_1 \theta_1 - \cos(j_1 + 1) \theta_1)(\cos j_2 \theta_2 - \cos(j_2 + 1) \theta_2) {d \, \theta_1 \over 4 \pi} {d\, \theta_2 \over 4 \pi}
}.
}
With $\Delta^a_b = \delta^a_b + \delta^{-a}_b$ the contribution from the factors in the denominator of \singlefactored\ is given by
\eqn\termone
{
I_1[d,j_1,j_2] = -\Delta^{j_1}_{j_2}\left(\Delta^d_{2 j_1} + \Delta^{d-4}_{2 j_1} + 2 \Delta^{d-2}_{2 j_1} \right) + \Delta^{d-1 \pm 1 }_{2 j_1}\Delta^{d-1 \pm 1 }_{2 j_2}
+ \Delta^{d-3 \pm 1 }_{2 j_1} \Delta^{d-3 \pm 1}_{2 j_2} - \Delta^{d-2 \pm 2}_{2 j_1}\Delta^{d-2 \pm 2}_{2 j_2}.
}
Next, we consider the $(1-x^2)$ factor in \singlefactored\ .
\eqn\secondterm{
-\log(1 - x^{2 k}) = \sum {x^{2 k n} \over n}.
}
This time, for the coefficient of $x^d$, we have
\eqn\littlesum{
\sum_{2 k | d} {2 \phi(k) \over d} = \left\{ \matrix{&1\quad &{\rm d \, even} \cr &0\quad &{\rm \, otherwise}} \right .
}
Substituting this into the main integral, we find that for $d>4$, we need to integrate
\eqn\secondint{
\matrix{\int 4 + 2(\cos \theta_1 + \cos \theta_2 + 1) dM & {\rm d \, even }, \cr \int -4 \cos {\theta_1 \over 2} \cos{\theta_2 \over 2} dM &{\rm d \, odd}.}
}
For $d < 4$ the expression above and below should be modified to drop terms that cannot contribute to the pole in $x$.

Define
\eqn\Itwodef
{
I_0[j_1,j_2] = 4 \Delta^0_{j_1}\Delta^0_{j_2} + \Delta^{j_1 - 1}_0 \Delta^{j_2}_0 + \Delta^{j_2 - 1}_{0}\Delta^{j_1}_0.
}
With $I = I_1 + I_0$, the contribution from the second term is
\eqn\termtwo{
I_2[d,j_1,j_2] = I[d,j_1,j_2] - I[d,j_1,j_2+1] - I[d,j_1 +1, j_2] + I[d,j_1+1,j_2+1].
}
Finally we consider the remaining quadratic term in \singlefactored\ . We will call $\sum_i q_i^k = 4 \cos{k \theta_1 \over 2} \cos{k \theta_2 \over 2}=\alpha_k$, to save space.
\eqn\quad{
-\log(1 - (\alpha_k x^k - x^{2 k})) = \sum_{n} {(\alpha_k x^k - x^{2 k})^n \over n} = \sum_{p,q}(-1)^q \alpha_k^p x^{(p + 2 q)k} {1 \over p + q} \pmatrix{p+q \cr q}.
}
Again we will want to collect the coefficient of $x^d$ here.  A term in the sum above contributes to this coefficient only if $p+ 2 q = d/k$. Also, this expression is
summed over $k$ against $\phi(k)/k$.  This means we need to consider the sum
\eqn\c{
\sum_{k|d}\sum_{p=0}^{d/k} {\phi(k)\over k}\alpha_k^p (-1)^{({d \over k} - p)/2} {2 \over {d\over k} + p} \pmatrix{{d \over 2 k} + p/2 \cr p}.
}
We now look at a generic integral, integrating this term against $\cos {A \theta_1 \over 2} \cos {B \theta_2 \over 2}$. All terms occurring in the actual integral of the
term in \quad\ may be reduced to this form.  Use the identity
\eqn\genint{
\int 4^p \left[\cos {k \theta_1 \over 2} \right]^p \cos {A \theta_1 \over 2} \left[ \cos{k \theta_2  \over 2} \right]^p \cos{B \theta_2 \over 2} =
\pmatrix{p \cr {1\over2}(p - {A \over k})} \pmatrix{p \cr {1 \over 2}(p - {B \over k})}.
}
To shorten expressions, define ${p \over 2} = s,  {d \over 2 k} = x, {A \over 2 k} = y,{B \over 2 k} = z$.

This allows us to write the generic integral over the sum in \c as
\eqn\isthisinteger
{S_1[d,A,B] =\sum_{k|(d,A,B)} \phi(k) \sum_{s = max(y,z)}^x (-1)^{x - s} \pmatrix{s+x \cr 2 s} \pmatrix{2 s \cr s-y} \pmatrix{2 s \cr s-z} {1 \over k (x + s)}.
}
Now define
\eqn\d{\eqalign{
& S_2[d,A,B]=S_1[d,A,B]+S_1[d-4,A,B]\cr
&-\sum_{\sigma_1,\sigma_2 = \pm 1/2}\bigg( S_1[d-1,A+\sigma_1,B+\sigma_2]+ S_1[d-3,A+\sigma_1,B+\sigma_2]\bigg ) \cr
&+ \sum_{\rho_1,\rho_2=\pm 1} S_1[d-2,A+\rho_1,B]+S_1[d-2,A,B+\rho_2]\cr
&+2 S_1[d-2,A,B].}
}
We can now collect all the terms that appear in the integral over the quadratic term \quad
\eqn\e{
I_3[d,j_1,j_2]= S_2[d,j_1,j_2]+ S_2[d,j_1+1,j_2+1]-S_2[d,j_1+1,j_2]-S_2[d,j_1,j_2+1].
}
Collecting the terms contributing, one finds the following enlightening result:
\eqn\finalanswer{
\int dM Z[\theta,\theta_1,\theta_2] \bar{\chi}^*_{[d,j_1,j_2]}= I_2[d,j_1,j_2]+I_3[d,j_1,j_2].
}
where $I_2$ is defined in \termtwo\ and $I_3$ in \e\ .
As we noted above, for $d < 4$ the expressions get modified.

These sums are prohibitively difficult to evaluate by hand, but may be easily done with a computer.
A {\it Mathematica} script which evaluates the sum above, with all special cases included, is available online at {\it http://people.fas.harvard.edu/$\sim$llhuang/conformalresult.nb}. 

It is easy to list the operators in the theory at low scaling dimension:

\bigbreak
\bigskip
{
\offinterlineskip
\def\tablerule{\noalign{\hrule}}
\def\tableskip{\omit&height 3pt&&&\omit\cr}
\halign{\tabskip = .7em plus 1em
\vtop{\hsize=6pc\pretolerance = 10000\hbadness = 10000
\normalbaselines\noindent\it#\strut}
&\vrule #&#\hfil &\vrule #
&\vtop{\hsize=11pc \parindent=0pt \normalbaselineskip=2pt
\normalbaselines \rightskip=3pt plus2em #}\cr
\noalign{\hrule height2pt depth2pt \vskip3pt}
\multispan5\bf Low Dimension Operators in the Pure YM Theory\hfil\strut\cr
\noalign{\vskip3pt} \tablerule
\omit&height 3pt&\omit&&\omit\cr
\bf Dimension&&\bf Operators&&\omit \bf Conformal Representation \hfil\cr
\noalign{\vskip -2pt}
&& &&\omit \bf Content \hfil\cr
\tableskip 1 &&no operators&& no representations\cr
\tableskip 2 &&$F_{\mu \nu}$&&
[2,1,0]+[2,0,1]\cr
\tableskip 3 &&no primary operators
&&no representations\cr
\tableskip 4&&$F_{\mu \nu} *F^{\mu \nu}$, $|F|^2$
&&[4,0,0]+[4,0,0]\cr
\tableskip &&$F_{\{ \mu \nu}F^{\nu}_{\sigma \}}-|F|^2$&&[4,1,1] \cr
\tableskip &&$F_{\{\mu \nu}F_{\rho \sigma \}}-
F_{\{ \mu \nu}F^{\nu}_{\sigma \}}-|F|^2$&&[4,2,0]+[4,0,2]\cr
\tableskip \tablerule \noalign{\vskip 2pt} \tablerule
}}

\bigskip

Now consider large values of $d$. Neglecting the angular variables, we see that, for large values of $d$ 
\eqn\asymptot{
Z = -\sum {\phi(k) \over k} \ln {(1+x)(x^2 - 4 x + 1) \over (1 - x)^3} \approx \beta^d x^d . 
}
where $\beta = 2 + \sqrt 3$ is the larger root of the quadratic term. This is the characteristic Hagedorn growth in the number of states. It is easy to verify, numerically, that \finalanswer does reproduce the right growth in the density of states.
 
As a final consistency check, it is necessary for \finalanswer to sum to an integer for every value of $d,j_1,j_2$; this is not at all apparent from the expression we have. Nevertheless, using some elementary number-theoretic results, we show, in the appendix, that our answer always sums to an integer.

We list here the representation content of the theory up to dimension 9.

\bigskip
{
\offinterlineskip
\def\tablerule{\noalign{\hrule}}
\def\tableskip{\omit&height 3pt&&\omit\cr}
\halign{\tabskip = .7em plus 1em
\vtop{\hsize=6pc\pretolerance = 10000\hbadness = 10000
\normalbaselines\noindent\it#\strut}
&\vrule #&#\hfil &\vrule #
&\vtop{\hsize=11pc \parindent=0pt \normalbaselineskip=2pt
\normalbaselines \rightskip=3pt plus2em #}\cr
\noalign{\hrule height2pt depth2pt \vskip3pt}
\multispan5\bf Representation Content of Pure YM \hfil\strut\cr
\noalign{\vskip3pt} \tablerule
\omit&height 3pt&\omit&&\omit\cr
\bf Dimension&&\bf Representations \hfil\cr
\noalign{\vskip -2pt}
\tableskip 2 &&[2,0,1] + [2,1,0]\cr
\tableskip 3 && nothing\cr
\tableskip 4 && 2[4,0,0] + [4,0,2] + [4,2,0] + [4,1,1]\cr
\tableskip 5 &&[5,3/2,3/2]\cr
\tableskip 6 &&2[6,0,0] + 2[6,0,1] + [6,0,3] + 2 [6,1,0] + 2[6,1,1]\cr
\tableskip   &&+ [6,1,2] + [6,1,3] + [6,2,1] + [6,2,2] + [6,3,0] + [6,3,1]\cr
\tableskip 7 && 4[7,1/2,3/2] + 2[7,1/2,5/2] + 4[7,3/2,1/2] + 4[7,3/2,3/2]\cr
\tableskip   &&+ 2[7,3/2,5/2] + 2[7,5/2,1/2]  +  2 [7,5/2,3/2] + [7,5/2,5/2]\cr
\tableskip 8 &&6[8,0,0] + 4[8,0,1] + 5[8,0,2] + [8,0,3] + 2 [8,0,4]\cr
\tableskip   &&+ 4 [8,1,0]+ 10 [8,1,1] + 7[8,1,2] + 5[8,1,3] + [8,1,4]\cr
\tableskip   &&+ 5[8,2,0] + 7 [8,2,1]+ 8 [8,2,2] + 3[8,2,3] + [8,2,4]\cr
\tableskip   &&+ [8,3,0]+ 5[8,3,1] + 3[8,3,2] + [8,3,3]+ 2[8,4,0]\cr
\tableskip   &&+ [8,4,1] + [8,4,2]\cr
\tableskip 9 &&14[9,1/2,1/2] + 20[9,1/2,3/2]+  15[9,1/2,5/2] + 6[9,1/2,7/2]\cr
\tableskip   && + 20[9,3/2,1/2] + 28[9,3/2,3/2]  + 18[9,3/2,5/2] + 7[9,3/2,7/2]\cr
\tableskip   &&+ 2 [9,3/2,9/2] + 15[9,5/2,1/2] +  18 [9,5/2,3/2] + 12[9,5/2,5/2]\cr
\tableskip   &&+ 4 [9,5/2,7/2] + 6[9,7/2,1/2] + 7 [9,7/2,3/2] + 4 [9,7/2,5/2]\cr
\tableskip   &&+ [9,7/2,7/2]+ 2[9,9/2,3/2] \cr
\tableskip \tablerule \noalign{\vskip 2pt} \tablerule
}}
\newsec{$\CN = 4$ SYM}

In principle, it is not difficult to generalize the procedure above to the case of the $\CN = 4$ Yang Mills Theory. This theory has an exact superconformal symmetry.  Representations of the Superconformal group are labeled by the highest weight under $SO(4) \times SO(2)$ and the {\it R charges}. These were originally classified in \DobrevQV. They were studied in \DobrevVH, \DobrevQZ, \DobrevSP and are discussed in detail in \DolanZH .

It is easy to generalize the partition function by adding in chemical potentials for the R charges. This result can be read off from the appendix in \DolanZH. Similarly, it is simple to generalize the result for the Haar measure and the characters\DobrevTK.

Unfortunately, short representations in the superconformal case have a rather more intricate structure than in the conformal case and it is not always possible
to write them as a difference of two long representations. This complication is not important in the spectrum of single trace operators in the $\CN=4$ theory, because it is known that the only relevant short representations are the ${1 \over 2}, {1 \over 2}$ BPS multiplets.

The more serious complication is numerical. Performing the character decomposition involves finding the coefficient of a specified monomial in a given power series expansion. Since we have six chemical potentials in the supersymmetric case, the simple algorithms are $O(d^6)$. Thus, the calculation quickly becomes unfeasible.

In a set of papers \refs{\BianchiWX, \BeisertTE, \BeisertDI, \BianchiWW, \BianchiXI}, Bianchi et. al. conjectured that the spectrum of the free SYM theory may be obtained from the spectrum of type IIB theory on flat space through a specified algorithm. They verified their conjecture using a sieve procedure which allowed them to determine the spectrum up to scaling dimension 10.

Further verification of this conjecture must await either a deeper understanding of their result or the development of more efficient numerical techniques.

\newsec{Conclusions}
Unitary representations of the Conformal Algebra must obey $d \geq j_1 + j_2 + 2$, for $j_1 j_2 \neq 0$ and $d \geq j_1 + j_2 + 1$ otherwise. Depending on whether either of these bounds is saturated, the characters of the conformal group fall into three classes. These are described in \characters.

The Free Yang Mills theory on a sphere has an exact conformal symmetry. Hence, its partition function may be written as a sum over the characters above. Formally, we have the result
\eqn\symbolicresult
{
Z = \sum N_{d,j_1,j_2} \chi_{d,j_1,j_2}.
}
In this note, we performed this decomposition. We find that $N_{d,j_1,j_2}$ is described by \finalanswer. Our formula demonstrates the correct asymptotics. Moreover, it is possible to prove that it always produces an integer.

\bigskip

\centerline{\bf Acknowledgements} The authors would like to thank Shiraz Minwalla for suggesting this problem and providing guidance.  We would like to thank Joseph Marsano, Andrew Neitzke and Xi Yin for several useful discussions. LLH would like to thank the Tata Institute of Fundamental Research, for its hospitality, while this work was being completed. The work of LLH was supported in part by an NSF Graduate Research Fellowship.
This work was supported in part by DOE grant DE-FG03-91ER40654 and by the NSF career grant PHY-0239626.
\listrefs

\newsec{Appendix}
Consider \isthisinteger, which has a sum over $k$ and
$s$. Schematically, the set of allowed $k,s,x$ values is shown below. In the figure, for each value of $k$, $s$ can range over the values demarcated by the horizontal line at the bottom and the outermost line. 

The critical point is to partition this large set of values correctly. We group the set of $k,s$ values in subsets of the kind $S_{x_0,s_0} = \{s,k,x,y,z: x + s = p(x_0 + s_0)\}$ This foliation is indicated on the diagram.

\fig{Grouping the Terms}
{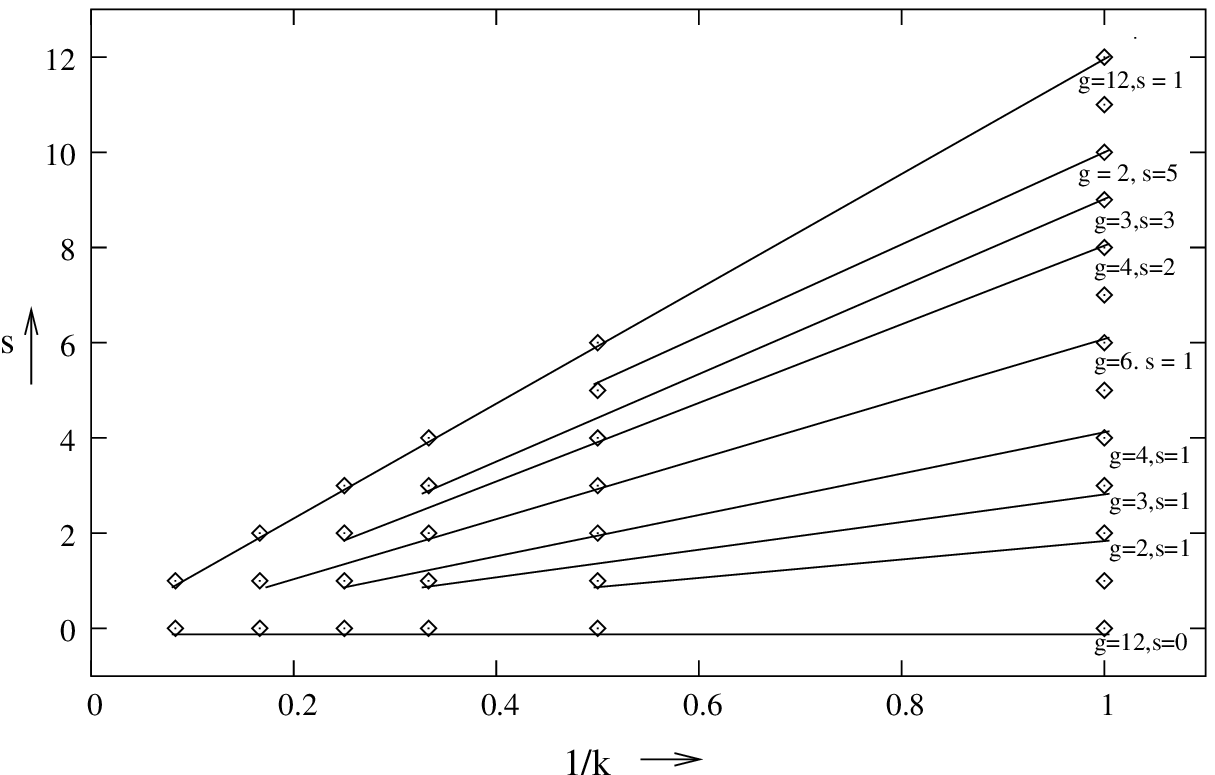}{3.75truein}
\figlabel{\groupfig}

It is clear that in each partition, we have $p k = g $, where $g$ is a constant.
Second, we have ${\rm gcd}(s_0 + x_0,x_0 - s_0, s_0 - y_0, s_0 - z_0, s_0) = 1$.
Since \isthisinteger\ can be written as:
$$
{\alpha_1 \over g(x_0 + s_0)} = {\alpha_2 \over 2 g s_0} = {\alpha_3 \over g(x_0 - s_0)} = {\alpha_4 \over g(s_0 - y_0)} ...
$$
where the $\alpha_i$ are integral, it suffices to show that the sum is divisible by g to show that it is an integer.

This leaves us to prove the following statement:
\eqn\isthisinteger
{\sum_{k|g} \phi(g/k)  (-1)^{k (x_0 - s_0)} \pmatrix{(s_0+x_0) k \cr 2 k s_0} \pmatrix{2 k s_0 \cr (s_0-y_0)k} \pmatrix{2 k s_0 \cr k(s_0-z_0)} {\rm mod} g = 0.
}
For notational simplicity, we consider the simpler statement,
\eqn\lemma{ \sum_{k | g}(-1)^{n_3 k} \pmatrix{n_1 k \cr n_2 k} \phi(g/k) \mod g = 0,}
where  $n_i$ are arbitrary integers with ${\rm gcd} \, {n_i} = 1$. Furthermore, we take $g = p^t$, where $p$ is prime. We take $p \neq 2$ so that $(-1)^{n_3 k}$ has the same sign. The generalization of this proof to generic $g$ is straightforward.

With $t = 1$, our sum is:
\eqn\lemmatone
{\pmatrix{n_1 \cr n_2} (t - 1) + \pmatrix {n_1 t \cr n_2 t} = 0 (\mod t).}
where we have used $\pmatrix {n_1 t \cr n_2 t} = \pmatrix {n_1 t \cr n_2 t} \mod t.$
Assume the statement is true for $t = n$. For $t = n+1$, our sum is:
\eqn\lemmainduction
{
\sum_{i=0}^{n+1} \pmatrix{n_1 p^i \cr n_2 p^i} \phi(p^{n+1-i}) =p  \sum_{i = 0}^{n} \pmatrix{n_1 p^i \cr n_2 p^i} + \pmatrix{n_1 p^{n+1} \cr n_2 p^{n+1}} - \pmatrix{n_1 p^n \cr n_2 p^n}
.}
The first term is divisible by $p^{n+1}$ by hypothesis. With $n_3 = n_1 - n_2$, write the second term as:
$$
\pmatrix{n_1 p^n \cr n_2 p^n} \left( {n_1 p^{n+1} (n_1 p^{n+1}-1) ... (n_1 p^n +1)- n_2 p^{n+1}  ... (n_2 p^n + 1) n_3 p^{n+1}  ... (n_3 p^n + 1) \over n_2 p^{n+1} ... n_2 p^n n_3 p^{n+1} .. n_3 p^n } \right).
$$
We can cancel leading terms divisible by $p^{n+1}$ in the numerator and denominator, but then we notice that subleading terms not divisible by $p^{n+1}$ cancel in the numerator but not in the denominator. So the second term is also divisible by $p^{n+1}$.
This proves our result.

\end{document}